\documentclass[10pt,sigconf,screen,authorversion]{acmart}

\usepackage[all]{nowidow}
\usepackage{xcolor}
\usepackage{acronym}
\usepackage{subcaption}
\usepackage{hyperref}

\usepackage[capitalise,noabbrev]{cleveref}
\crefformat{section}{\S#2#1#3}
\crefformat{subsection}{\S#2#1#3}
\crefformat{subsubsection}{\S#2#1#3}

\begin{document}

\author{Tobias Pfandzelter}
\affiliation{%
    \institution{TU Berlin \& ECDF}
    \city{Berlin}
    \country{Germany}}
\email{tp@mcc.tu-berlin.de}

\author{David Bermbach}
\affiliation{%
    \institution{TU Berlin \& ECDF}
    \city{Berlin}
    \country{Germany}}
\email{db@mcc.tu-berlin.de}

\title{Edge Computing in Low-Earth Orbit -- What Could Possibly Go Wrong?}



\acmYear{2023}\copyrightyear{2023}
\setcopyright{acmlicensed}
\acmConference[LEO-NET '23]{Workshop on Low Earth Orbit Networking and Communication}{October 6, 2023}{Madrid, Spain}
\acmBooktitle{Workshop on Low Earth Orbit Networking and Communication (LEO-NET '23), October 6, 2023, Madrid, Spain}
\acmPrice{15.00}
\acmDOI{10.1145/3614204.3616106}
\acmISBN{979-8-4007-0332-4/23/10}

\begin{abstract}
    Large low-Earth orbit (LEO) satellite networks are being built to provide low-latency broadband Internet access to a global subscriber base.
    In addition to network transmissions, researchers have proposed embedding compute resources in satellites to support LEO edge computing.
    To make software systems ready for the LEO edge, they need to be adapted for its unique execution environment, e.g., to support handovers in face of satellite mobility.

    So far, research around LEO edge software systems has focused on the predictable behavior of satellite networks, such as orbital movements.
    Additionally, we must also consider failure patterns, e.g., effects of radiation on compute hardware in space.
    In this paper, we present a taxonomy of failures that may occur in LEO edge computing and how they could affect software systems.
    From there, we derive considerations for LEO edge software systems and lay out avenues for future work.
\end{abstract}

\maketitle

\section{Introduction}
\label{sec:intro}

SpaceX, Amazon Kuiper, OneWeb, and other private aerospace companies are leveraging new technologies such as reusable rockets and free-space laser communication to build massive communication constellations in low-Earth orbit (LEO).
Hundreds or thousands of satellites per constellation, orbiting at altitudes of 500 to 2,000km, will provide global low-latency broadband Internet access.~\cite{Pultarova2015-ml,Handley2018-ay,Bhattacherjee2018-vc}.

In addition to communication, researchers have suggested augmenting satellites with compute resources, enabling \emph{LEO edge computing}~\cite{Bhosale2020-aa,Bhattacherjee2020-kr,paper_pfandzelter_LEO_serverless,paper_pfandzelter2022celestial}.
By embedding application services in the network, improved QoS for emerging application areas can be achieved, and network link strain can be alleviated with local data processing.

Deploying applications on satellite servers at the LEO edge is not trivial:
Satellite mobility, limited compute capabilities, energy budgets, geographical distribution of clients, and the constraints of orbital mechanics all control how these servers can be used.
Researchers have suggested abstracting from these challenges with compute platforms that automate client handover and data movement~\cite{Bhattacherjee2020-kr,Bhosale2020-aa,paper_pfandzelter_LEO_serverless}.

Due to the lack of operational LEO edge infrastructure, the research community is only able to investigate \emph{predictable} behaviors of satellite servers, e.g., imposed by orbital mechanics.
In addition, a key part of prototyping software for the LEO edge is an understanding of possible \emph{ad-hoc} characteristics of satellite server infrastructure, i.e., \emph{faults}~\cite{bagchi2019dependability}.

In this paper, we investigate what failure patterns exist in LEO networks and satellite servers and how they may impact LEO edge software systems.
We give an overview of satellite communication constellations (\cref{sec:background}) and existing studies of their real world performance (\cref{sec:relwork}).
In \cref{sec:failures}, we then present a taxonomy of possible failures from a software system perspective, including ground-to-satellite communication degradation, compute component faults in space, and effects of orbital safety maneuvers.
We review network performance studies, astronomical and physical models, and regulatory filings; study the behavior of existing LEO communications satellites based on public ephemerides; model radiation effects; and perform simulations of LEO satellite constellations.\footnote{We make all artifacts used to produce this paper available at \url{https://github.com/pfandzelter/leo-edge-failure-models}.}
Finally, we propose future research directions on making LEO edge software resilient \cref{sec:resiliency}.

\section{Background}
\label{sec:background}

\begin{figure}
    \centering
    \includegraphics[width=1\linewidth]{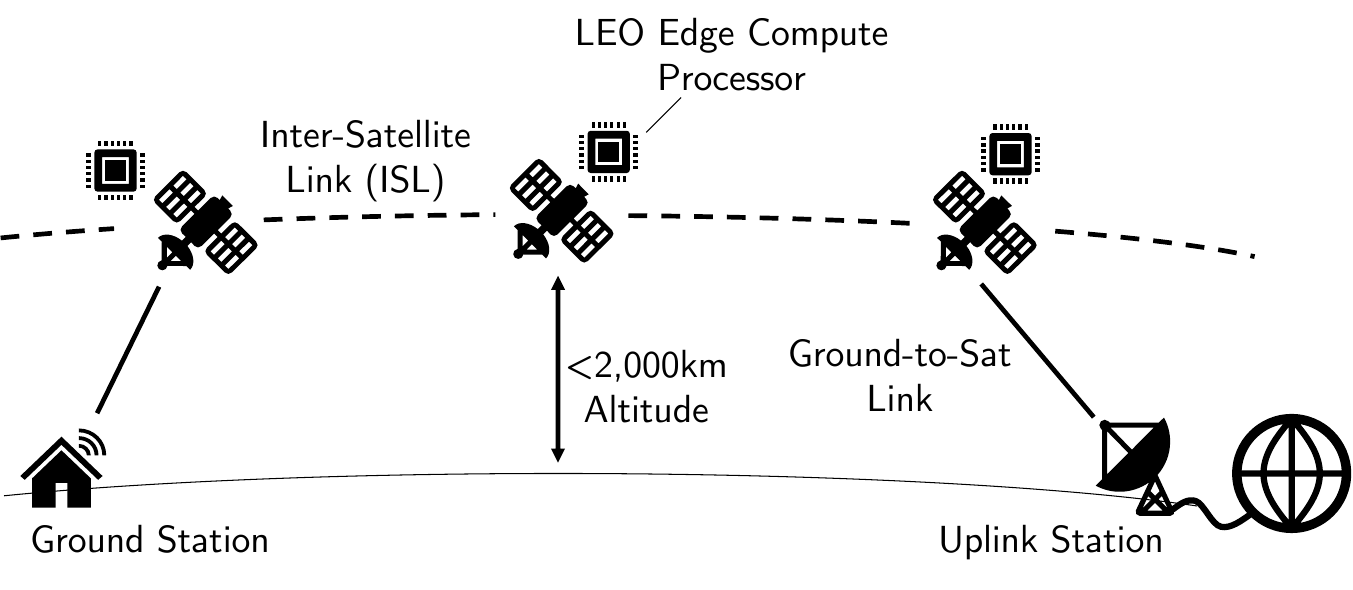}
    \caption{Components of a LEO Satellite Communications Network: Satellites in LEO relay network communications between ground stations and uplink stations using inter-satellite links (ISL). Additional processing components on satellites enable LEO edge computing.}
    \Description{A radio antenna on a house that is connected to a satellite in low-Earth orbit. Two more satellites are connected using inter-satellite links. All satellites are equipped with compute hardware. Satellite altitude is shown to be less than 2,000km above Earth. A large antenna on Earth is the uplink station. This uplink station is connected to a satellite using a ground-to-satellite link. The antenna is also connected to the Internet.}
    \label{fig:leo-architecture}
\end{figure}

Satellite-based Internet access has been available for decades, using communications relay satellites in geostationary orbit at altitudes of more than 35,000km above Earth, where satellites appear at a fixed point in the sky relative to an observer on Earth.
While this makes ground signals easy to direct, the high communication delays (550ms round-trip) at this altitude make it a solution only for niche use-cases~\cite{10.1145/3517745.3561432}.

With technological advances in phased array antennas, free-space optical links, and satellite launch and development, LEO satellite constellations are now a viable alternative for global broadband access.
At altitudes of less than 2,000km above Earth, communication delays are on the order of only 20ms~\cite{michel2022first}.
Satellite mobility and smaller cones of coverage makes constellations of satellites necessary, e.g., the \emph{Gen1} Starlink constellation encompasses 4,408 satellites, with more than 3,000 already operational~\cite{jonathansspacepage}.

We show an overview of a such a communication constellation in \cref{fig:leo-architecture}.
Ground stations, which can be radio antennas on homes, planes, or even mobile phones~\cite{tmobilestarlink2022}, connect to satellites over radio links.
These satellites are connected among each other using free-space optical inter-satellite links (ISL) to cover greater distances with high bandwidth and low latency.
An uplink ground station is connected to the Internet and acts as the subscriber's access point.

In addition to Internet access, researchers have proposed to embed compute resources on satellites for in-network processing~\cite{Bhattacherjee2020-kr,Bhosale2020-aa,paper_pfandzelter_LEO_serverless}.
As an extension of terrestrial multi-access edge computing (MEC), LEO edge computing, where data can be processed on network paths, supports latency and privacy goals while addressing bandwidth limitations~\cite{shi2016promise}.

\section{Related Work}
\label{sec:relwork}

The concept of LEO edge computing was outlined by Bhattacherjee et al.~\cite{Bhattacherjee2020-kr}.
The authors describe use-cases and replication techniques to counteract orbital movements of satellites but leave fault-tolerance to future work.
Compute platforms for LEO edge software systems~\cite{Bhosale2020-aa,paper_pfandzelter_LEO_serverless} as well as simulation~\cite{Kassing2020-yc,paper_pfandzelter_LEO_CDN,Kempton2021-lw} and emulation~\cite{paper_pfandzelter2022celestial,techreport_pfandzelter_celestial_extended,kon2022stargaze} tools have been proposed.
These tools and platforms, however, only focus on the predictable behavior of satellite constellations, such as those derived from orbital trajectories of satellites.

As argued by Singla~\cite{singla2021satnetlab}, there is little public information on the actual characteristics of satellite networks, let alone that of the LEO edge.
Kassem et al.~\cite{kassem2022browser} perform a browser-based study of the Starlink service using volunteer nodes in Europe, Australia, and the USA.
They find that it can provide better connectivity than terrestrial networks they test, yet bandwidth and network delays vary between subscriber locations.
Further, their results show an impact of weather and satellite handovers on packet loss.
Michel et al.~\cite{michel2022first} and Ma et al.~\cite{ma2022network} have conducted similar studies of Starlink.

As no LEO edge resources are deployed in current commercial satellite constellations, measurement studies from the client side can unfortunately not yield knowledge about faults of compute resources in space.
Wang et al.~\cite{wang2021tiansuan} have proposed the \emph{Tiansuan Constellation} of six research satellites in LEO equipped with CPU and GPU resources to perform software testing in space.
While this will support evaluation of LEO edge research under the most realistic conditions, it is expensive and thus limited in scale.
To the best of our knowledge, resiliency challenges in LEO edge computing have not yet been studied holistically.
\section{Failure Models}
\label{sec:failures}

Considering the existing research, we identify five main points of possible failures in a LEO communications constellation with edge computing capabilities:
(i) degradation of up- and downlinks (\cref{sec:failures:links}), (ii) failures of on-board compute hardware (\cref{sec:failures:hardware}), (iii) deviation from orbital paths of a satellite (\cref{sec:failures:orbital}), (iv) degradation of ISLs (\cref{sec:failures:isl}), and (v) adversarial attacks against LEO constellations (\cref{sec:failures:adversaries}).

\subsection{Up-/Downlink Degradation}
\label{sec:failures:links}

\begin{figure}
    \centering
    \begin{subfigure}{0.475\linewidth}
        \centering
        \includegraphics[width=1\linewidth]{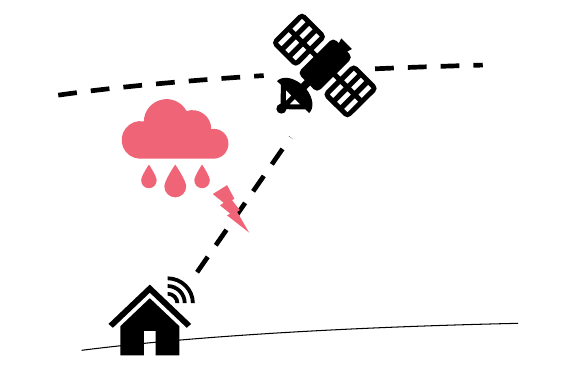}
        \caption{Rain and other weather can decrease RF link signal strength, impacting observed network throughput.}
        \label{fig:linkdegradation:rain-fade}
        \Description{A house with a radio antenna is connected to a satellite in low-Earth orbit using RF links. A red rain cloud disturbs the link, illustrating effects of rain on ground-to-satellite radio links.}
    \end{subfigure}%
    \hfill
    \begin{subfigure}{0.475\linewidth}
        \centering
        \includegraphics[width=1\linewidth]{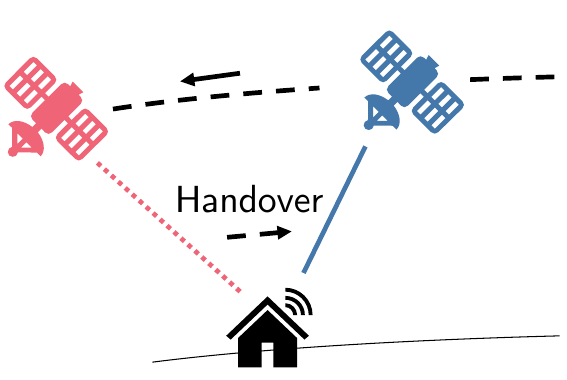}
        \caption{Orbital movements of LEO satellites lead to frequent handovers, interrupting uplink connections.}
        \label{fig:linkdegradation:handover}
        \Description{A house with a radio antenna is connected to a red satellite in low-Earth orbit with a radio link. The red satellite is shown to move to the left. A blue satellite is shown that moves from the right to the left. The ground connection is handed over from the red to the blue satellite.}
    \end{subfigure}%
    \caption{The two main causes for ground-to-sat link degradation in LEO satellite networks are rain fade (\cref{fig:linkdegradation:rain-fade}) and handovers between satellites (\cref{fig:linkdegradation:handover}).}
    \label{fig:linkdegradation}
\end{figure}

Communication satellites use radio links to communicate with ground stations on Earth.
As illustrated in \cref{fig:linkdegradation}, two effects can occasionally decrease link performance, i.e., reducing available bandwidth for subscribers and increasing packet loss:
rain attenuation (\cref{fig:linkdegradation:rain-fade}) and satellite handovers (\cref{fig:linkdegradation:handover}).

Starlink relies on the K\textsubscript{a} and K\textsubscript{u} band for up- and downlinks~\cite{fccfiling2022starlinkgen2}, which are subject to atmospheric attenuation.
Vashisht et al.~\cite{vasisht2021l2d2} have shown that signal strength for the K\textsubscript{a} band decreases even with clouds.
Real-world measurements with Starlink, however, show a less pronounced effect:
We find no correlation between the client-side Starlink performance reported by Michel et al.~\cite{michel2022first} and historical weather data for their user terminal location, where it only rained lightly (<2mm per hour) during their measurement period.
Kassem et al.~\cite{kassem2022browser} find effects on browser-side Starlink performance only with moderate rain, with network wait times doubling.
Similarly, Ma et al.~\cite{ma2022network} observe a 50\% drop in download throughput (215Mb/s to 120Mb/s) with moderate rainfall (>4mm precipitation).
Note that rain at both \emph{client} and \emph{gateway} locations can affect observed throughput.

As LEO imposes a significant speed on satellites, continuous network connection also requires frequent handovers.
Kassem et al.~\cite{kassem2022browser} observe spikes of 1-2\% packet loss rates every 1-2 minutes, which likely correlate with handovers and regularly impact the network performance.

\subsection{On-Board Compute Failure}
\label{sec:failures:hardware}

Compute hardware in LEO is subject to increased effects of radiation compared to sea level.
Protons and atomic nuclei originating from outside Earth that hit electronics with sufficient energy cause signal spikes and noise.
Especially problematic are trapped particles in the inner Van Allen belt stretching between altitudes of 640km and 9,600km~\cite{nasa2014vanallen} and as low as 200km in the South Atlantic Anomaly~\cite{stassinopoulos2015forty}.

The results are single-event upsets (SEU) and hardware degradation due to total ionizing dose (TID).
SEU are caused by a single particle leading to functional interrupts (SEFI) and latch-ups (SEL).
SEFI are soft errors that can interrupt correct program execution or cause the operating system or hardware to fail.
With soft errors, nominal execution can be achieved with reset of the system, while SEL can permanently damage individual components.

TID are the long-term, cumulative effects of radiation that lead to permanent equipment malfunction.
TID thus effectively determines the lifetime of a component~\cite{maurer2008td}.

Satellite flight-control hardware usually uses expensive, low-performance radiation-hardened components with performance that lacks behind the state of the art.
Commercial off-the-shelf (COTS) compute hardware is more feasible for LEO edge compute as it is more efficient and applications are not mission-critical.
With the \emph{Spaceborne Computers}~\cite{noauthor_2021-jm}, Hewlett Packard Enterprise has installed such COTS hardware on the ISS without issues, although the ISS itself is radiation-shielded to protect human occupants.
Similarly, the \emph{OrbitsEdge SatFrame\texttrademark}~\cite{orbitsedge2022} uses shielding for COTS components, decreasing SEU rates and ionizing dose.
Tiansuan constellation servers will use clusters of small COTS processors~\cite{wang2021tiansuan}, where each SEU only affects a part of the system.

\begin{figure}
    \centering
    \includegraphics[width=1\linewidth]{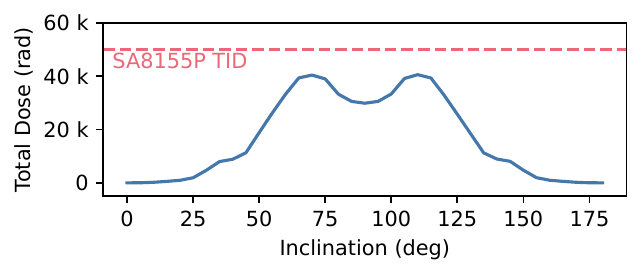}
    \caption{Mission total ionizing dose for circular orbits at 550km altitude and different inclinations assuming a 1mm aluminum shielding. The dashed red line marks the TID limit of a Snapdragon SA8155P SoC~\cite{sheldon2022radiation}.}
    \Description{A graph showing orbital inclination between 0 and 180 degrees on the x-axis and total dose between 0 and 50krad on the y-axis. A red dashed line at 50krad shows the TID limit for the Snapdragon SA8155P SoC. A blue line shows mission TID for different inclinations, mirrored at 90 degrees. The mission TID is largest at 73 degrees at about 40krad and smallest at 0 degrees at about 0rad.}
    \label{fig:totaldose}
\end{figure}

Hardware and firmware can be designed to catch radiation-induced errors, but recovery without system resets is unlikely~\cite{noauthor_2021-jm}.
We argue that from a software and operating system perspective, these faults will appear as unscheduled system reboots.
The Tiansuan in-space computing server~\cite{wang2021tiansuan} will likely use a cluster of 60 Qualcomm \emph{Snapdragon 865} SoCs (cf.~Zhang et al.~\cite{zhang2022soc}).
Radiation testing results for the \emph{Snapdragon SA8155P}, which has a similar architecture, put its TID at $\sim$50krad for nominal execution~\cite{sheldon2022radiation}.
Using a 550km altitude circular LEO orbit with different inclinations, we calculate the dose rates using the \emph{AP8-MIN}~\cite{vette1991ae}, \emph{AE8-MAX}~\cite{sawyer1976trapped}, and \emph{SHIELDOSE-2}~\cite{seltzer1994updated} models part of ESA's \emph{Space Environment Information System}~\cite{kruglanski2009space}.
As shown in \cref{fig:totaldose}, 1mm aluminum shielding is sufficient to protect an SoC for a five-year lifetime~\cite{Bhattacherjee2020-kr,spacexfiling2022b}.
For reference, a sufficient aluminum box for the HPE ProLiant DL325 Gen10 used in the feasibility analysis by Bhattacherjee et al.~\cite{Bhattacherjee2020-kr} would increase the weight of the server by 10.8\% and its volume by 5.49\%.

For the SoC server with this shielding, we estimate a SEU-induced soft error rate on the order of $10^{-3}$ to $10^{-4}$ per device per day based on the SEFI/SEL measurements of Sheldon et al.~\cite{sheldon2022radiation} and extrapolations with the \emph{CREME-96} model~\cite{tylka1997creme96}.
We note, however, that this is exacerbated with the number of devices:
If the \emph{Gen1} Starlink constellation were equipped with similar compute resources (60 SoCs per each of the 4,408 satellites), between 30 and 300 SEU-induced errors would occur every 24 hours.

\subsection{Orbital Maneuvers}
\label{sec:failures:orbital}

\begin{figure}
    \centering
    \begin{subfigure}{0.475\linewidth}
        \centering
        \includegraphics[width=1\linewidth]{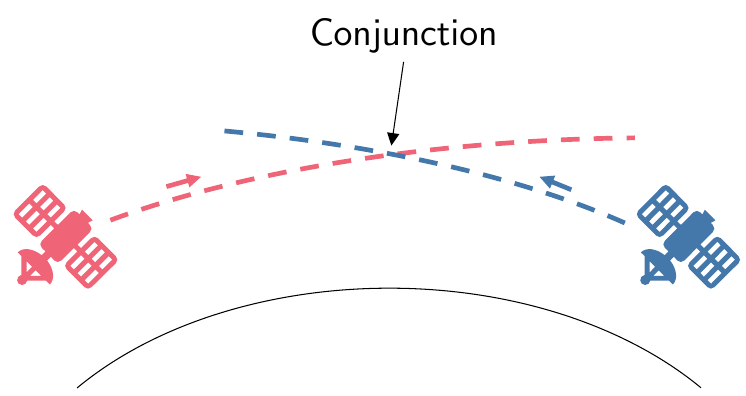}
        \caption{High-Risk Conjunction}
        \label{fig:conjunction:risk}
        \Description{A red and a blue satellite in low-Earth orbit are shown. The satellites follow red and blue trajectory lines, respectively. The lines are shown to interset, and the word conjunction marks this intersection.}
    \end{subfigure}%
    \hfill
    \begin{subfigure}{0.475\linewidth}
        \centering
        \includegraphics[width=1\linewidth]{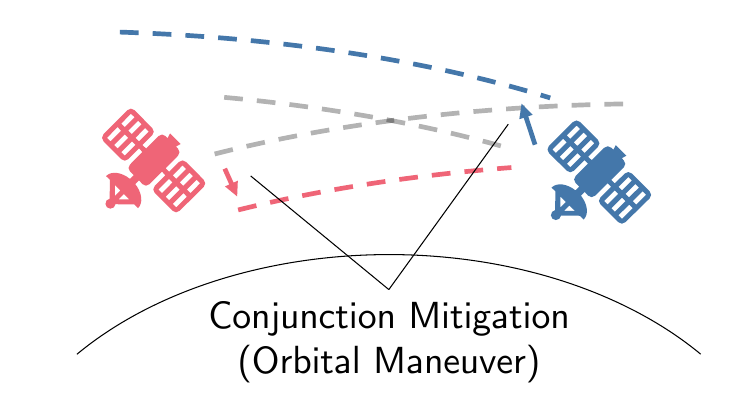}
        \caption{Conjunction Mitigation}
        \label{fig:conjunction:avoidance}
        \Description{A conjunction mitigation using an orbital maneuver is illustrated for a red and a blue satellite. The red satellite lowers its orbit, while the blue satellite increases its altitude.}
    \end{subfigure}%
    \caption{The crowded LEO leads to frequent conjunctions between objects, e.g., satellites, debris, space stations. High-risk conjunctions can be identified in frequent simulations (\cref{fig:conjunction:risk}). Orbital maneuvers must be performed to reduce the risk of the conjunctions, changing satellite trajectories (\cref{fig:conjunction:avoidance}).}
    \label{fig:conjunction}
\end{figure}

In theory, satellites remain in the position induced by the parameters of their constellation for the entirety of their lifetime.
Through controlled maneuvers to counteract orbital decay, e.g., caused by atmospheric drag, we expect satellites to follow strictly Keplerian orbits, which makes it easier to predict networking and compute characteristics~\cite{dang2022spaceweather,fang2022spaceweather}.

Nevertheless, orbital debris and other satellites in LEO necessitate frequent orbital adjustments to avoid high-risk conjunctions, i.e., possible collisions in space.
As shown in \cref{fig:conjunction:risk}, such a conjunction can occur if the orbital trajectories of two objects intersect with a high (e.g., greater than 1 in 10,000~\cite{satelliteorbitalsafety2022,spacexfiling2022}) probability of collision (Pc).
This risk is usually identified by official and commercial conjunction assessment providers based on tracking data~\cite{krage2020nasa}.

Using on-board thrusters, such as the krypton-powered Hall effect thrusters used in Starlink satellites~\cite{starlink2019presskit}, high-risk conjunctions can be avoided by adjusting orbits, as shown in \cref{fig:conjunction:avoidance}.
In total, Starlink satellites have performed 13,612 such propulsive maneuvers between June and November 2022, with each satellite averaging 12 maneuvers per year~\cite{spacexfiling2022}.
With an increasing number of objects in LEO, we can expect a higher frequency of conjunction mitigation in the future: SpaceX budgets their second-generation Starlink satellites for 350 maneuvers over their five-year lifetime~\cite{spacexfiling2022b}.

Despite their frequent occurrence, we find the impact of such maneuvers on the LEO network to be limited.
We compare two-line element set~\cite{kelso1998tle} and \emph{SOCRATES} satellite orbital conjunction reports~\cite{kelso2005satellite} for Starlink satellites for two weeks in January 2023.
Our results show that for satellites at high risk of conjunction ($\text{Pc} \ge {10}^{-4}$), only minor orbital adjustments are made, with only a 1-3km change in altitude.
This altitude change is a controlled maneuver that lowers conjunction risk to acceptable levels but is unlikely to have significant effects on the LEO network characteristics:
First, a controlled maneuver allows operators to anticipate ISL orientation to prevent link interruption between satellites.
Second, the small link length change is insignificant for network delay variation on the ISL.

\subsection{ISL Degradation}
\label{sec:failures:isl}

\begin{figure}
    \centering
    \includegraphics[width=1\linewidth]{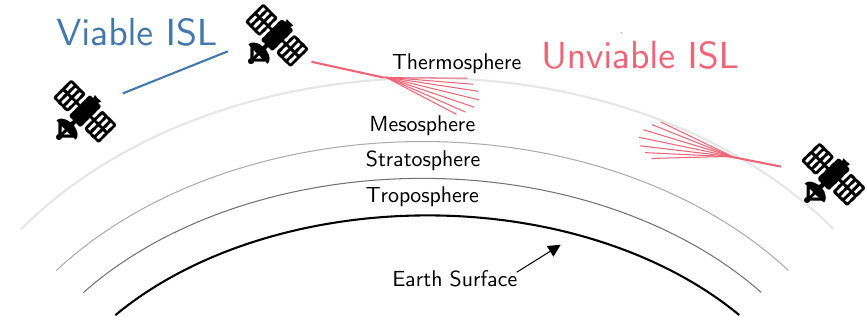}
    \caption{Free-space laser ISLs only work above the lower Earth atmosphere (above ~80km), as particles can refract the laser.}
    \label{fig:islrefraction}
    \Description{Three satellites in low-Earth orbit are shown. The Earth surface is marked. Above that, in increasing altitude, Troposphere, Stratosphere, Mesosphere, and Thermosphere are marked. The two left satellites are connected with a blue link that says Viable ISL. This link is above the Thermosphere. The two right satellites are connected with a broken red line that says Unviable ISL. This line goes through the Thermosphere.}
\end{figure}

\begin{figure}
    \centering
    \includegraphics[width=1\linewidth]{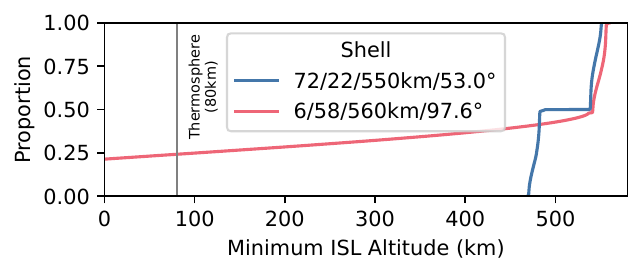}
    \caption{A one-hour simulation of ISLs in two \emph{Gen1} Starlink shells shows that only a fraction of laser links pass through the atmosphere below altitudes of 80km. The minimum link altitude depends on the spacing between orbital planes, with most Starlink, Kuiper, and OneWeb links well above 400km.}
    \label{fig:linkheights}
    \Description{A graph showing a cumulative distribution for minimum ISL altitudes is marked. The x-axis shows minimum altitude in kilometers between 0 and 600. The y-axis shows proportion between 0 and 1. The Thermosphere is marked with a gray vertical line at 80km. A blue line for a Starlink shell with 72 satellites on 22 orbital planes at 550km altitude and 53 degrees orbital inclinations shows the lowest altitudes of ISLs. These are 50\% in the range of about 450km to 500km and 50\% in the range of 500 to 550km. A blue line for a Starlink shell with 6 satellites on 58 orbital planes at 560km altitude and 97.6 degree orbital inclination shows the lowest altitude of ISLs. These are 50\% at 550km but 50\% lower than 550km, with about 25\% infeasible as they cross below the Thermosphere.}
\end{figure}

ISLs are important to provide the low round-trip times of satellite networks and for communication between services in LEO edge computing.
Laser ISLs in LEO are unlikely to be affected by obstruction or weather attenuation, yet atmospheric refraction of the laser may lead to connection interruptions.
Although successful free-space laser communication between satellites with the lowest altitude of the line of sight at 20km have been achieved~\cite{smutny20095}, typical estimates for a minimum safe altitude are $\sim$80km~\cite{Bhattacherjee2019-jz}.
As shown in \cref{fig:islrefraction}, this is the height of the Mesosphere, where water vapors and other gases can attenuate the laser links and cause refraction~\cite{Bhattacherjee2019-jz}.
Although even a link below this altitude could work with decreased signal strength, current constellations rarely use ISLs in this area given their orbital parameters.

As shown in \cref{fig:linkheights}, only the two sparsest \emph{Gen1} Starlink shells have ISLs that are blocked by the Earth, other shells and constellations are not subject to ISL degradation.
Note that we assume a \emph{+GRID} topology where a satellite connects to its neighbors in its own orbital plane as well as one satellite in each adjacent plane over ISLs.
Link setup delays make other, dynamic topologies infeasible~\cite{Bhattacherjee2019-jz,Bhattacharjee2023-wf}.
This minimum ISL altitude can be easily anticipated in network simulation and taken into account in routing.

\subsection{Adversarial Attacks}
\label{sec:failures:adversaries}

Finally, satellite networks are exposed to adversarial attacks.
In addition to typical software vulnerabilities that may come with running arbitrary software on shared infrastructure, network links can be disrupted, e.g., by jamming ground-to-satellite links, orbits can be changed, and satellites can even be destroyed completely~\cite{howell2022starlinkhacked,blinken2021antisatellite,Pavur2020-wf,donsez2022cubedate}.
Unlike the previous challenges, which are induced by the unique environment of LEO and physical and engineering constraints, adversarial attacks are entirely unpredictable.
Offering LEO edge infrastructure as-a-Service means trusting third parties to run arbitrary software on satellites, requiring strong isolation from mission-critical systems and other tenants.

\section{Research Opportunities}
\label{sec:resiliency}

The failure scenarios we have reviewed in \cref{sec:failures} show that LEO edge software systems are subject to unique fault scenarios.
In this section, we give an overview of research opportunities in resilience that address some of these scenarios.

\subsubsection*{Dynamic Communication Strategies}

Existing research on LEO edge computing assumes perfect network conditions, e.g., a predictable network topology and link bandwidth, for their communication strategies~\cite{paper_pfandzelter2022celestial,Bhattacherjee2019-jz}.
Ground-to-satellite link degradation and minor effects from orbital maneuvers, however, will impact network characteristics.
These variations are not currently taken into consideration in LEO edge  networking and communication strategies, but will need to be regarded in the future.
For example, rain fade can influence the bandwidth on the direct connection between the satellite network and a cloud data center, and alternative routes must be considered that could potentially influence the offloading strategy of a combined edge-cloud service.

\subsubsection*{Software Resiliency}

Even if on-board compute failures are unlikely, they cannot be ruled out completely.
Given the size of a LEO constellation, expensive recovery mechanisms for such failures will not be efficient.
Future work on LEO edge software, especially in the context of compute platforms~\cite{Bhosale2020-aa,paper_pfandzelter_LEO_serverless}, must also focus on making recovery from such failures fast and mostly transparent to application services.
A possible way forward is to decouple state, such as application-specific data, from compute services, e.g., using a serverless compute platform in combination with a database system~\cite{paper_pfandzelter_LEO_serverless}.

\subsubsection*{Compute \& Data Replication}

Radiation-induced errors and network variations also increase the need for data and compute service replication to provide a high QoS level for LEO edge services.
Efficient replication is increasingly complex with the system and network constraints of the LEO edge.
Designing and implementing replication strategies, as well as finding ways to efficiently fail-over from faulty satellite servers, will present an interesting area for future research.

\subsubsection*{Security}

Advances in securing LEO satellite networks~\cite{kon2022stargaze,giuliari2021icarus} mean that cyber-security threats in LEO networks can increasingly be mitigated.
Adversarial attacks on LEO edge software, however, must also deal with untrusted code in the constrained satellite server environment.
This will require novel research on efficient software isolation to prevent adversaries to influence other tenants' software, the LEO network, or even mission-critical flight hardware.

\section{Conclusion \& Future Work}
\label{sec:conclusion}

In this paper, we have presented an overview of failure models for LEO edge computing.
Link degradation, spontaneous network changes, on-board compute failures, and adversarial attacks can all influence the behavior of the LEO edge, and software systems targeting this environment must be able to cope with these failures.
Based on our review, we have suggested open research opportunities that need to be addressed to make LEO edge software resilient.
In future work, we plan to integrate our learnings into LEO edge simulation and emulation tools to quantify the impact on existing software and the feasibility of fault-tolerance techniques from terrestrial edge computing in this new environment.

\begin{acks}
    Funded by the \grantsponsor{DFG}{Deutsche Forschungsgemeinschaft (DFG, German Research Foundation)}{https://www.dfg.de/en/} -- \grantnum{DFG}{415899119}.
\end{acks}

\balance

\bibliographystyle{ACM-Reference-Format}
\bibliography{bibliography.bib}

\end{document}